\documentclass[
  ,draft            
]
{aipproc}

\usepackage{mathptm}
\usepackage{bm}
\usepackage{mtimes}

\layoutstyle{6x9}

\newcommand\beq{\begin{eqnarray}}
\newcommand\eeq{\end{eqnarray}}

\newcommand{\do}{^{\circ }}

\def\kvec{\vec{k}_\perp}
\def\k3vec{\mbox{\boldmath $k$}}
\def\bvec{\vec{b}}

\def\0vec{\mbox{\boldmath $0$}_\perp}

\def\vep{\varepsilon}

\def\slash#1{\rlap/{#1}}

\def\qslash{\slash{\mkern-1mu q}}

\def\vepslash{\slash{\mkern-1mu \varepsilon}}
\def\eslash{\slash{\mkern-1mu e}}
\def\Delslash{\slash{\mkern-1mu \Delta}}

\def\es{s}

\input{epsf.sty}
\input{epsfig.sty}
\input{amssymb.sty}

\begin{document}

\title{ Diffractive heavy pseudoscalar-meson
productions by weak neutral currents}

\author{A. Hayashigaki}{
  address={ Department of Physics, University of Tokyo, Tokyo 113-0033, 
Japan }
}

\author{ K. Suzuki }{
  address={ Division of Liberal Arts, Numazu College of Technology, Shizuoka 410-8501, Japan }
}

\author{ K. Tanaka }{
  address={ Department of Physics, Juntendo University, Inba-gun, 
Chiba 270-1695, Japan }
}

\begin{abstract}
A first theoretical study for neutrino-induced 
diffractive 
productions of heavy pseudoscalar-mesons, $\eta_c$ and $\eta_b$, off a nucleon
is performed based on factorization formalism in QCD.
We evaluate the forward diffractive production cross section 
in perturbative QCD in terms of the light-cone 
wave functions of 
$Z$ boson and 
$\eta_{c, b}$ mesons, and the gluon distribution of the nucleon.
The diffractive production of $\eta_c$
is governed by the axial vector coupling of the longitudinally polarized $Z$ boson
to $Q\bar{Q}$ pair, and the resulting $\eta_c$ production cross section  
is larger than the $J / \psi$ one
by one order of magnitude.
The bottomonium $\eta_b$ production, 
which shows up for higher beam energy, is also discussed. 
\end{abstract}

\maketitle


Exclusive diffractive leptoproductions of neutral vector mesons provide
unique insight into an interplay between nonperturbative and perturbative 
effects in QCD.
The diffractive processes are mediated 
by the exchange of a Pomeron with the vacuum quantum numbers,
whose QCD description is directly related to the gluon distributions inside the 
nucleons for small Bjorken-$x$ \cite{BFGMS,FKS,RRML}.
The processes also allow us 
to probe the light-cone wave functions (WFs)
of the vector mesons. Relating to the latter point, however, 
the applicability is 
apparently limited to probing the neutral vector mesons
due to the vector nature of the electromagnetic current.

In this talk, we propose the exclusive diffractive 
productions of mesons in terms of the 
neutrino beam. The weak currents
allow us to observe both neutral
and charged mesons by $Z$ and $W$ boson exchanges, and these mesons
can be not only vector but 
also other types of mesons including pseudoscalar mesons.
Thus, such processes may reveal structure of various kinds of mesons, 
the coupling of the QCD Pomeron to quark-antiquark pair with various spin-flavor
quantum numbers, and also
information on the CKM matrix elements.  
There already exist some experimental data for 
$\pi$, $\rho$, $D_s^{\pm}$, $D_s^{*}$ \cite{E815,E632}, 
$D_s^{*+}$ \cite{CHORUS}, and $J / \psi$ production \cite{E815}, although 
the amount of the data is not enough.
On the other hand, there are only a few theoretical calculations, e.g.,
for the $J / \psi$ production in a vector meson dominance model \cite{BKP}
and for $D_s^-$ production 
with the generalized parton density \cite{LS}. 
Here, our interest will be directed to diffractive 
productions of heavy pseudoscalar mesons, $\eta_c$ and $\eta_b$.
So far $\eta_c$ 
has been observed via the decays of $J/\psi$ or $B$ mesons
produced by $p\bar{p}$ and $e^+e^-$ reactions, while 
$\eta_b$ has not been observed.
The diffractive productions via the weak neutral current
will give a direct access to $\eta_{c}$
as well as a new experimental method to identify 
$\eta_b$
by e.g. measuring the two photon decay. 

\begin{figure}[h]
\psfig{file=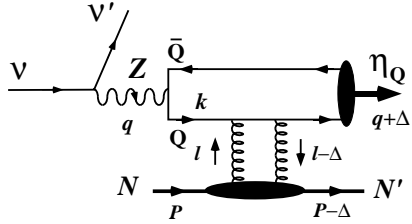,width=5.5cm}
\caption{
A typical diagram for the exclusive diffractive $\eta_Q$ ($Q=c,b$) productions 
induced by neutrino ($\nu$) through the $Z$ boson exchange. 
There are other diagrams by interchanging the vertices on the heavy-quark lines.}
\label{fig:etac_prod2}
\end{figure}

We treat the $\eta_c$ and $\eta_b$ productions by 
generalizing the approach in the leading logarithmic order of perturbative QCD, 
which has been developed 
successfully for the vector meson
electroproductions \cite{BFGMS,FKS,RRML}.
We consider the near-forward diffractive productions 
$Z^* (q)+ N (P) \to \eta_Q (q+\Delta) + N' (P-\Delta)$,
where $Q=c, b$, and each momentum is labeled in Fig.\ref{fig:etac_prod2}.
Here the total center-of-mass energy 
$W= \sqrt{(P + q)^2}$ is much larger than 
any other mass scales involved, i.e.,
$W^2 \gg 
{\cal Q}^{2}(=-q^2), -t(=-\Delta^2), m_{Q}^{2}, \Lambda_{\rm QCD}^2, \ldots$
with $m_{Q}$ the heavy-quark mass. $- t \ll m_{Q}^2$ and $m_Q^2\gg \Lambda_{\rm QCD}^2$ are also assumed.
The crucial point is 
that at such high $W$ the scattering of the $Q\bar{Q}$ pair on the nucleon
occurs over a much shorter timescale than the $Z^{*} \rightarrow Q\bar{Q}$
fluctuation or the $\eta_Q$ formation times (see Fig.\ref{fig:etac_prod2}).
As a result, the production amplitudes obey factorization 
in terms of the $Z$ and $\eta_{Q}$ light-cone WFs. Also, a hard scale $m_Q$ ensures the application of perturbative QCD 
for the $Z$ WFs even for ${\cal Q}^2 =0$.
The $Q \bar Q$-$N$ elastic scattering amplitude, 
sandwiched between the  $Z$ and $\eta_{Q}$ WFs,
is further factorized
into the $Q \bar Q$-gluon hard scattering amplitude 
and the nucleon matrix element corresponding to the (unintegrated) gluon density distribution \cite{BFGMS,FKS,RRML}.
The participation of the ``new players'' $Z$ and $\eta_Q$
requires an extension
of the previous works \cite{BFGMS,FKS,RRML}
by introducing the corresponding light-cone WFs.


First of all,
we discuss the extension due to the participation of the $Z$ boson.
The $ZQ\bar{Q}$ weak vertex of Fig.~\ref{fig:etac_prod2}
is given by 
$(g_W/2\cos \theta_W)\gamma_\mu (c_V-c_A\gamma_5)$,
where $(g_{W}/2 \cos \theta_W)^2 = \sqrt{2}G_F M_Z^2$ with $G_F$ the Fermi
constant and $M_Z$ the $Z$ mass. 
$c_V=1/2 - (4/3) \sin^2 \theta_W, c_A = 1/2$ for the $c$-quark and similarly 
for the $b$-quark.
As usual,
we introduce the two light-like vectors 
$q'$ and $p'$ by the relations
$q=q'-({\cal Q}^2/\es)p'$, $P=p' + (M_{N}^{2}/\es)q'$, $\es=2 q'\cdot p'$
with $M_N$ the nucleon mass,
and the Sudakov decomposition of all momenta, 
e.g., $k = \alpha q'+\beta p'+k_\perp$ ($k_\perp^2=-\kvec^2$).
We also introduce 
the polarization vectors $\vep^{(\xi)}$ ($\xi = 0, \pm 1$) of the virtual $Z$ boson to satisfy
$\sum_{\xi}(-1)^{\xi + 1}{\varepsilon_{\mu}^{(\xi)}}^{*}\varepsilon_{\nu}^{(\xi)}
= - g_{\mu \nu} + q_{\mu}q_{\nu}/M_{Z}^{2}$,
which is the numerator of the propagator for the massive 
vector boson. Because the 
$q_{\mu}q_{\nu}/M_{Z}^{2}$ term vanishes when contracted with the neutral current,
we conveniently choose as $\vep^{(0)} = q'/{\cal Q}+ p'{\cal Q}/\es$ and $\vep^{(\pm 1)} = \vep^{(\pm 1)}_{\perp} = (0,1,\pm i,0)/\sqrt{2}$
for the longitudinal ($\xi =0$) and transverse ($\xi = \pm 1$) polarizations respectively.
The light-cone WFs
for the virtual $Z$ boson can be obtained
in analogy with the photon light-cone WF used in the $J/\psi$ electroproductions \cite{BFGMS,FKS} as 
\begin{eqnarray}
\Psi^{Z(\xi)}_{\lambda \lambda'} (\alpha,\kvec) = - \frac{g_W}{2 \cos \theta_W}
\frac{\sqrt{\alpha(1-\alpha)} \bar{u}_{\lambda}(k) \vepslash^{(\xi)}
\left(c_{V} - c_{A}\gamma_{5}\right)v_{\lambda'}(q-k)}
{\alpha (1-\alpha) {\cal Q}^2 + \kvec^2 +m_Q^2},
\label{eqn:2-12}
\end{eqnarray}
where $u_{\lambda}(k)$ ($v_{\lambda'}(q-k)$) denote the on-shell spinor for the 
(anti)quark with helicity $\lambda^{(\prime)}$.


Next we proceed to the light-cone WFs for the $\eta_{Q}$ meson.
It is convenient to exploit the correspondence with
the case of the $J/\psi$ electroproductions.
The light-cone WFs for the heavy vector
mesons $V=J/\psi, \Upsilon$ have been discussed in many works,
but are still controversial 
in the treatment of subleading
effects like the Fermi motion corrections \cite{FKS,RRML,hood}, corrections to ensure
the pure $S$-wave $Q\bar{Q}$ state \cite{ivanov, HIKT}, etc.
Here we employ the vector-meson light-cone WFs
given by 
\begin{eqnarray}
{\Psi^{V(\varpi)}_{\lambda \lambda'}}^{*}(\alpha,\mbox{\boldmath $k$}_\perp)
&=&\frac{\bar{v}_{\lambda'}(q+\Delta-k)}{\sqrt{1-\alpha}}\gamma^{\mu}
{e_{\mu}^{(\varpi)}}^{*} {\cal R}
\frac{u_{\lambda}(k)}{\sqrt{\alpha}}
\frac{\phi^*(\alpha,\kvec)}{M_{V}} , 
\label{eqn:V}
\end{eqnarray}
where $M_{V}$ and $e_{\mu}^{(\varpi)}$ ($\varpi = 0, \pm 1$) are the mass 
and the polarization vector of the vector meson
with $(q+\Delta)^2 = M_{V}^{2}$, $e^{(\varpi)}\cdot (q+ \Delta) = 0$, and 
$e^{(\varpi) *}\cdot e^{(\varpi')}= - \delta_{\varpi \varpi'}$.
${\cal R} \equiv \left[1+(\qslash+\Delslash)/M_{V}\right]/2$ denotes the projection operator,
${\cal R}^2 = {\cal R}$,   
to ensure the $S$-wave $Q\bar{Q}$ state in the heavy-quark limit \cite{hood,BJ}.
(This projection operator coincides with that discussed in Ref.~\cite{ivanov}
up to the binding-energy effects of the quarkonia.)
Note that eq.~(\ref{eqn:V}) reduces to the vector-meson WFs of Ref.~\cite{BFGMS}
by the replacement ${\cal R} \rightarrow 1$.
Essential difference of eq.~(\ref{eqn:V}) from the ``perturbative'' WFs (\ref{eqn:2-12}) is that the scalar function $\phi(\alpha,\kvec)$ contains 
nonperturbative dynamics between $Q$ and $\bar{Q}$. 
Now the light-cone WFs of the $\eta_{Q}$ meson 
can be derived from eq.~(\ref{eqn:V}) utilizing spin symmetry,
which is exact in the heavy-quark limit. 
This symmetry relates the $S$-wave states, $\eta_{Q}$ and
the three spin states of the vector meson.
Namely, $M_{\eta_Q} = M_{V}$, and 
the pseudoscalar state is related to the vector state with
longitudinal polarization as
$|\eta_{Q} \rangle = 2 \hat{S}_{Q}^{3} |V(\varpi=0)\rangle$,
where 
$\hat{S}_{Q}^{3}$ is the third component of
the hermitean spin operator $\hat{S}_{Q}^{i}$ 
which acts on the spin of the heavy quark $Q$ but does not act on $\bar{Q}$.
This implies that the $\eta_Q$ WFs are given by the replacement 
$u_\lambda \rightarrow 2 S^{3} u_\lambda$ in eq.~(\ref{eqn:V}), 
where $S^{3}$ is a matrix representation of $\hat{S}_{Q}^{3}$
as $S^{3} = \gamma_{5} (\qslash+\Delslash) \eslash^{(0)}/(2M_{V})$
which is related to a spin matrix $\sigma^{12}/2=\gamma_{5}\gamma^{0}\gamma^{3}/2$
in the meson rest frame
by a Lorentz boost in the third direction:
\begin{eqnarray}
{\Psi^{\eta_{Q}}_{\lambda \lambda'}}^{*}(\alpha,\mbox{\boldmath $k$}_\perp)
=-\frac{\bar{v}_{\lambda'}(q+\Delta -k)}{\sqrt{1-\alpha}}\gamma_5
{\cal R}
\frac{u_{\lambda}(k)}{\sqrt{\alpha}} \ 
\frac{\phi^*(\alpha,\kvec)} {M_{\eta_Q}} .
\label{eqn:3-1}
\end{eqnarray}
This result shows that the $\eta_{Q}$ is described by the same 
nonperturbative WF
$\phi(\alpha,\kvec)$ as the vector meson.
We also note that, due to the presence of ${\cal R}$,
the ``$Q\bar{Q}\eta_{Q}$ vertex'' involves pseudovector 
as well as pseudoscalar coupling.

Combining our $Z$ and $\eta_{Q}$ WFs 
with the $Q\bar{Q}$-$N$ elastic amplitude \cite{BFGMS},
we get the total amplitude ${\cal M}^{(\xi)}$ for the polarization $\varepsilon^{(\xi)}$ of the virtual $Z$ boson. We find ${\cal M}^{(\pm 1)} = 0$, which reflects conservation of helicity
in the high energy limit, and 
\begin{eqnarray}
i{\cal M}^{(0)}
&=&
\frac{-\sqrt{2}\pi^2 W^2}{\sqrt{N_c}}\frac{g_W m_Q c_A}{M_{\eta_Q}{\cal Q}\cos \theta_W}
\alpha_s({\cal Q}_{\rm eff}^2)\!
\left[1+i\frac{\pi}{2}\frac{\partial}{\partial 
\mbox{ln}x}\right]
\nonumber\\
&\times& xG(x,{\cal Q}_{\rm eff}^2)\int_0^1 
\frac{d\alpha\tilde{\cal Q}}{\alpha(1-\alpha)}
\int_0^\infty  dbb^2\phi^*(\alpha,b)
K_1\left(b\tilde{\cal Q}\right) ,
\label{eqn:5-1}
\end{eqnarray}
where $x=({\cal Q}^2+M_{\eta_Q}^2)/\es$, ${\cal Q}^2_{\rm eff}=({\cal Q}^2+M_{\eta_Q}^2)/4$, 
$\tilde{\cal Q}=[\alpha(1-\alpha){\cal Q}^2+m_Q^2 ]^{1/2}$, 
$K_{1}$ is a modified Bessel function, and
$G(x, {\cal Q}_{\rm eff}^{2})$ is the conventional gluon distribution. Eq.~(\ref{eqn:5-1}) is written in the ``$\bvec$-space'' conjugate to the $\kvec$-space via the 
Fourier transformation;
$\bvec$ denotes the transverse separation ($b\equiv |\bvec|$) between $Q$ and $\bar{Q}$. In derivation of eq.~(\ref{eqn:5-1}) we retain only the leading $\ln ({\cal Q_{\rm eff}}^2/\Lambda_{\rm QCD}^{2})$ contribution, which corresponds to the ``color-dipole picture'' \cite{FKS,SHIAH}.
As expected, the result (\ref{eqn:5-1}) is proportional to $c_A$
so that the $\eta_{Q}$ meson is generated by the axial-vector part of the 
weak current.
(For comparison, we also calculate the diffractive 
vector meson production via the weak neutral current. 
Using eq.~(\ref{eqn:V}), 
we find that ${\cal M}^{(0)}$
and ${\cal M}^{(\pm 1)}$ give the production of the longitudinally and transversely
polarized vector mesons, respectively,
and that all these amplitudes
are proportional to the vector coupling $c_{V}$.)

Combining eq.~(\ref{eqn:5-1}) with the $Z$ boson propagator and the weak neutral current by a neutrino, 
the forward differential cross section for the $\eta_{Q}$ production 
is given by
\begin{eqnarray}
\left.\frac{d^3\sigma(\nu N \rightarrow \nu'N' \eta_Q )}{d\es d{\cal Q}^2 dt}
\right|_{t=0}
=&&
\!\!\!\!
\frac{1}{4(8\pi)^3E_\nu^2 M_N^{2} \es}
\frac{g_W^2}{\cos \theta_W^2}
\frac{{\cal Q}^2}{\left({\cal Q}^2+M_Z^2\right)^2}
\frac{\epsilon}{1-\epsilon}
\left|{\cal M}^{(0)}
\right|^2,
\label{eqn:5-2}
\end{eqnarray}
where $E_{\nu}$ is the neutrino beam energy in the lab system, and 
$\epsilon = [4(1-y)-Q^2/E_\nu^2]/[2\{1+(1-y)^2\}+Q^2/E_\nu^2]$ 
with $y=\es /(2M_{N}E_\nu)$.
In order to evaluate 
the corresponding elastic $\eta_{Q}$ production rate,
we assume the $t$-dependence as 
$d^3\sigma/d\es d{\cal Q}^{2} dt =d^{3}\sigma/d\es d{\cal Q}^{2} dt|_{t=0}\, \exp (B_{\eta_{Q}} t)$ 
with a constant diffractive slope, as in the case of the vector meson production.
Integrating over $t$, ${\cal Q}^{2}$ and $\es$,
we get the elastic production rate
$\sigma(\nu N \rightarrow \nu'N' \eta_Q )$
as a function of $E_\nu$.

\begin{figure}[htb]
\psfig{file=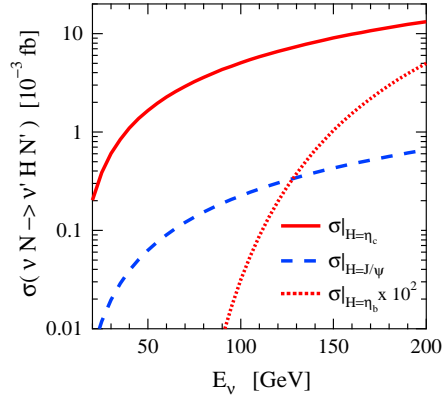,width=6.3cm}
\caption{The elastic production rates as functions of $E_\nu$.
Solid, dashed and dotted curves are for $\eta_c$, $J /\psi$ and $\eta_{b}$,
respectively. Note that the dotted curve shows the rate multiplied 
by $10^2$.}
\label{fig:tot_etac}
\end{figure}
For numerical computation of $\sigma( \nu N \rightarrow \nu'N' \eta_Q )$,
we need explicit form of the nonperturbative WF $\phi(\alpha,\bvec)$ of
eq.~(\ref{eqn:5-1}). As mentioned above,
we can use the corresponding nonperturbative part of the vector-meson WF
and we use the one which was constructed in Ref.~\cite{SHIAH} based on
non-relativistic Cornell potential model for heavy quarkonia with $m_{c(b)}=1.5(4.9)$ GeV.
Also, we use the empirical values for the masses $M_{\eta_c}=2.98$ GeV, 
$M_N=0.94$ GeV and $M_Z=91.2$ GeV. For $\eta_{b}$, we use an estimate $M_{\eta_b}=9.45$ GeV \cite{HK}. 
Because the slope $B_{\eta_{Q}}$ introduced above is unknown, 
we assume that $B_{\eta_{Q}}$ has the same value
as that for the corresponding vector meson:
$B_{\eta_{c}} =4.5$ GeV$^{-2}$ and $B_{\eta_b}=3.9$ GeV$^{-2}$ 
(See Ref.~ \cite{RRML}).
For the gluon distribution function $G(x,{\cal Q}_{\rm eff}^2)$ 
of eq.~(\ref{eqn:5-1}),
we employ 
GRV95 NLO parameterization \cite{GRV95}.

We show the elastic $\eta_{c}$ production rate,
$\sigma( \nu N \rightarrow \nu'N' \eta_c )$,
by the solid curve in Fig.~\ref{fig:tot_etac}.
The result monotonically increases as a function of the beam energy $E_\nu$.
Such behavior is similar with that observed in the $\pi$-production data 
by the neutral and charged currents \cite{E632}.
For comparison, we show the  
elastic $J/\psi$ production rate $\sigma( \nu N \rightarrow \nu'N' J/\psi )$
by the dashed curve.
The rate for $\eta_c$ production 
is much larger than that for $J/\psi$
by a factor $\sim 20$.
This is mainly due to the relevant weak couplings,
$c_{A}$ for $\eta_{c}$ and $c_{V}$ for $J/\psi$, 
as $(c_{A}/c_{V})^{2} \cong 7$. Also, most of the remaining factor arises from the different behavior of the $Z$ light-cone WFs (\ref{eqn:2-12}) 
between the axial-vector and vector channels,
resulting in a few times difference in the overlap integrals
with the corresponding meson WFs.
In Fig.~\ref{fig:tot_etac}, we also show the $\eta_{b}$ production rate
$\sigma( \nu N \rightarrow \nu'N' \eta_b )$
by the dotted curve.
Although the rate for $\eta_{b}$ is generally much
smaller than that for $\eta_{c}$,
the former increases more rapidly than the latter
for increasing $E_\nu$.
Therefore, the $\eta_{b}$ production rate
could become comparable with the $J/\psi$ or $\eta_c$ productions
for higher beam energy. It suggests a possibility
to observe $\eta_{b}$ through the diffractive productions 
by high intensity neutrino beams available at ongoing or forthcoming
neutrino factories.

In conclusion, we have computed the diffractive production cross sections of 
$\eta_{c}$ and $\eta_{b}$ mesons
via the weak neutral current. 
Using the new results of the light-cone WFs for $Z$ and $\eta_{c,b}$,
the production rates are obtained based on the factorization formalism in QCD.
Our results demonstrate that neutrino-induced productions will open
a new window to measure $\eta_{c,b}$.

\begin{theacknowledgments}
A.H. was supported by JSPS Research Fellowship for Young Scientists.
\end{theacknowledgments}




\bibliography{sample}

\begin{thebibliography}{99}
\bibitem{BFGMS} S.J. Brodsky {\it et al.},
\ Phys. Rev. {\bf D50}, 3134 (1994).

\bibitem{FKS} L. Frankfurt {\it et al.},
Phys. Rev. {\bf D57}, 512 (1998).

\bibitem{RRML} M.G.Ryskin {\it et al.},
\ Z. Phys. {\bf C76}, 231 (1997).

\bibitem{E815} E815 Collaboration, T. Adams {\it et al.},\ Phys. Rev. {\bf 
D61}, 092001 (2000).

\bibitem{E632} E632 Collaboration, S. Willocq {\it et al.},\ Phys. Rev. 
{\bf D47}, 2661 (1993).

\bibitem{CHORUS} CHORUS Collaboration, P. Annis {\it et al.},\ Phys. Lett. 
{\bf B435}, 458 (1998).

\bibitem{BKP} 
J.H. K$\ddot{\mbox{u}}$hn and R. R$\ddot{\mbox{u}}$ckl,\ Phys. Lett. {\bf B95}, 431 (1980).

\bibitem{LS} B. Lehmann-Dronke and A. Sch$\ddot{\mbox{a}}$fer,
Phys. Lett. {\bf B521}, 55 (2001).

\bibitem{hood}
P.~Hoodbhoy,
Phys. Rev. {\bf D56}, 388 (1997).

\bibitem{ivanov}
I.~P.~Ivanov and N.~N.~Nikolaev,
JETP Lett. {\bf 69}, 294 (1999).


\bibitem{HIKT} J. H\"{u}fner {\it et al.},
\ Phys. Rev. {\bf D62}, 094022 (2000).

\bibitem{BJ}
E.~L.~Berger and D.~L.~Jones,
Phys. Rev. {\bf D23}, 1521 (1981).

\bibitem{SHIAH} K. Suzuki {\it et al.},
\ Phys. Rev. {\bf D62}, 031501(R) (2000).

\bibitem{HK} D.S. Hwang and G.-H. Kim,\ Z. Phys. {\bf C76}, 107 (1997).

\bibitem{GRV95} M. Gl$\ddot{\mbox{u}}$ck {\it et al.},
Z. Phys. {\bf C67},  433 (1995).

\end{thebibliography}

\IfFileExists{\jobname.bbl}{}
 {\typeout{}
  \typeout{******************************************}
  \typeout{** Please run "bibtex \jobname" to optain}
  \typeout{** the bibliography and then re-run LaTeX}
  \typeout{** twice to fix the references!}
  \typeout{******************************************}
  \typeout{}
 }

\end{document}